\documentclass{appolb}
\usepackage{epsfig}

\begin{document}
\title{D meson nuclear modification factors in Pb-Pb collisions at $\sqrt{s_{NN}}$ = 2.76 TeV measured with the ALICE detector at the CERN-LHC
\thanks{Stangness in Quark Matter SQM2011 -- Cracow}%
}
\author{Alessandro Grelli on behalf of the ALICE Collaboration
\address{ERC- Research Group QGP-ALICE, Utrecht University, Princeton plein 5, \\3584 CC Utrecht, The Netherlands}
}
\maketitle
\begin{abstract}
The properties of the hot and dense QCD medium formed in ultra-relativistic heavy ion collisions, as well as the mechanism of in-medium partonic energy loss, can be accessed via the study of the D mesons nuclear modification factor.  
The ALICE experiment has measured D meson production in pp and Pb-Pb collisions at the LHC at $\sqrt{s}$ =7  and 2.76 TeV and at $\sqrt{s_{NN}}$  = 2.76 TeV, respectively, via the exclusive reconstruction of hadronic decay channels. 
D mesons are selected by exploiting the high-resolution tracking performance and the hadron identification capabilities of the ALICE detectors.
In this contribution we report on the analyses of the D$^0\rightarrow $K$^-\pi^+$, the D$^+ \rightarrow $K$^- \pi^+ \pi^+$  and the D$^{*+}\rightarrow $D$^0 \pi^+$  channels.  The preliminary results on D mesons nuclear modification factors are presented.
\end{abstract}

\section{Introduction}

Under conditions of high energy density and temperature, produced in ultra-relativistic high energy heavy-ion collisions, lattice QCD calculations predicts the transition from ordinary matter to a de-confined state of quarks and gluons called Quark Gluon Plasma (QGP)~\cite{w}.  Heavy-flavor quarks, differently from light quarks and gluons, are produced at the early stage of the collision in high-virtuality scattering processes. They traverse the medium and are expected to be sensitive to its density through the mechanism of in-medium partonic energy loss. Heavy flavours should loose less energy than light-quarks and  gluons as a consequence of a mass-dependent restriction in the phase space into which gluon can occur~\cite{eloss3}.
The nuclear modification factor ($\rm R_{AA}$) of D mesons, obtained by comparing their production in proton-proton and heavy ion collisions, allows to probe the properties of the formed high density QDC medium and the mechanism of in-medium energy loss. The $R_{AA}$ is defined as:
\begin{equation}
R^{{\rm D}}_{{\rm AA}}(p_{\rm t}) =\frac{1}{<\rm T_{{\rm AA}}>} \frac{{\rm d}N^{\rm D}_{{\rm AA}}/ \rm d p_{\rm t}}{\rm d\sigma^{\rm D}_{{\rm pp}}/\rm d p_{\rm t}}
\end{equation}
 where $N^{\rm D}_{{\rm AA}}$ is the yield in A--A collisions, $<\rm T_{{\rm AA}}>$, in a given centrality class, is the average nuclear overlap function calculated via Glauber model and $\sigma^{D}_{pp}$ is the production cross section of D mesons in pp collisions. In absence of medium effects $R^D_{AA}$  is expected to be 1.
In this contribution the evolution of the nuclear modification factor of D$^0$, D$^+$ and D$^{*+}$ versus transverse momentum is presented in the centrality ranges $0-20\%$ and $40-80\%$. A comparison with $R^{\pi^+}_{AA}$ and with shadowing~\cite{def} calculations is made.

\section{Detector and data sample}
  
In the following the relevant detectors for the D meson reconstruction are discussed; a detailed description of the ALICE apparatus is available in reference~\cite{aliceJINST}. 
The central barrel detectors are contained in a large solenoid magnet, which provides a field of 0.5~T. 
The closest detector to the beam axis is the Inner Tracking System (ITS). It is made of six cylindrical layers of silicon detectors with radii between 3.9 and 43.0~cm. The total material budget for radial tracks in the transverse plane amounts to $\sim$ 7.7$\%$ of the radiation length~\cite{Dmes}. 
Its location close to the interaction point together with the low magnetic field of the experiment allow the ITS to track low $p_t$ hadrons and to improve the momentum resolution of the ALICE tracking system. 
The ITS is surrounded by the main ALICE tracking detector, a 510~cm long cylindrical Time Projection Chamber (TPC). It provides track reconstruction with up to 159 three-dimensional space-points as well as particle identification via specific energy deposit $dE/dx$. The Time-Of-Flight detector (TOF), is positioned in the region 377 to 399~cm from the beam axis. The TOF complements the hadron identification capability of the TPC, ensuring an efficient track by track kaon/pion separation up to a momentum of about 1.5 GeV/$c$. With the present level of calibration, the intrinsic timing resolution is better than 100 ps.

\subsection{Trigger conditions}

The analyses presented in this paper are based on the Pb--Pb data sample at centre-of-mass energy $\sqrt{s_{NN}}=2.76$ TeV
collected in November 2010 during the first run with heavy-ions at the LHC.
The events were collected with a minimum-bias trigger based on the information 
of the Silicon Pixel Detector (SPD, $|\eta|<2$) and the VZERO scintillator hodoscopes  ($2.8<\eta<5.1$ and $-3.7<\eta<-1.7$) .
The efficiency for triggering hadronic interactions was 100$\%$ for Pb--Pb collisions in the centrality range considered in the analyses.
Beam background collisions were removed offline on the basis of the timing information provided by the VZERO and the 
neutron ZDC detectors (located near the beam pipe at $z\pm 114$~m from the interaction point).
Only events with a vertex found within 10~cm from the centre of the detector along the beam line were used, for a total of $17\times 10^6$ collisions.

\subsection{Centrality definition}
The Pb--Pb collisions  were classified based on their centrality defined in terms of  percentiles of the hadronic Pb--Pb cross section and determined from the
distribution of the summed amplitudes in the VZERO scintillator hodoscopes. This distribution was fitted using the Glauber model for the geometrical 
description of the nuclear collision~\cite{glauber} together with a two-component  model for particle production~\cite{ALICE-PbPbMult1}.

\section{D meson reconstruction}

The D$^{0}$, D$^+$ and D$^{*+}$ mesons were reconstructed in their hadronic channels: D$^0 \rightarrow $K$^- \pi^+$ (BR = ($3.91\pm0.05)\%$), D$^+ \rightarrow $K$^+\pi^-\pi^+$ (BR = $(9.22\pm 0.21)\%$) and D$^{*+} \rightarrow $D$^0\pi^+$ (BR = ($67.7\pm 0.5)\%$). 
\begin{figure}[h!]
\begin{center}
\includegraphics[width=0.32\textwidth]{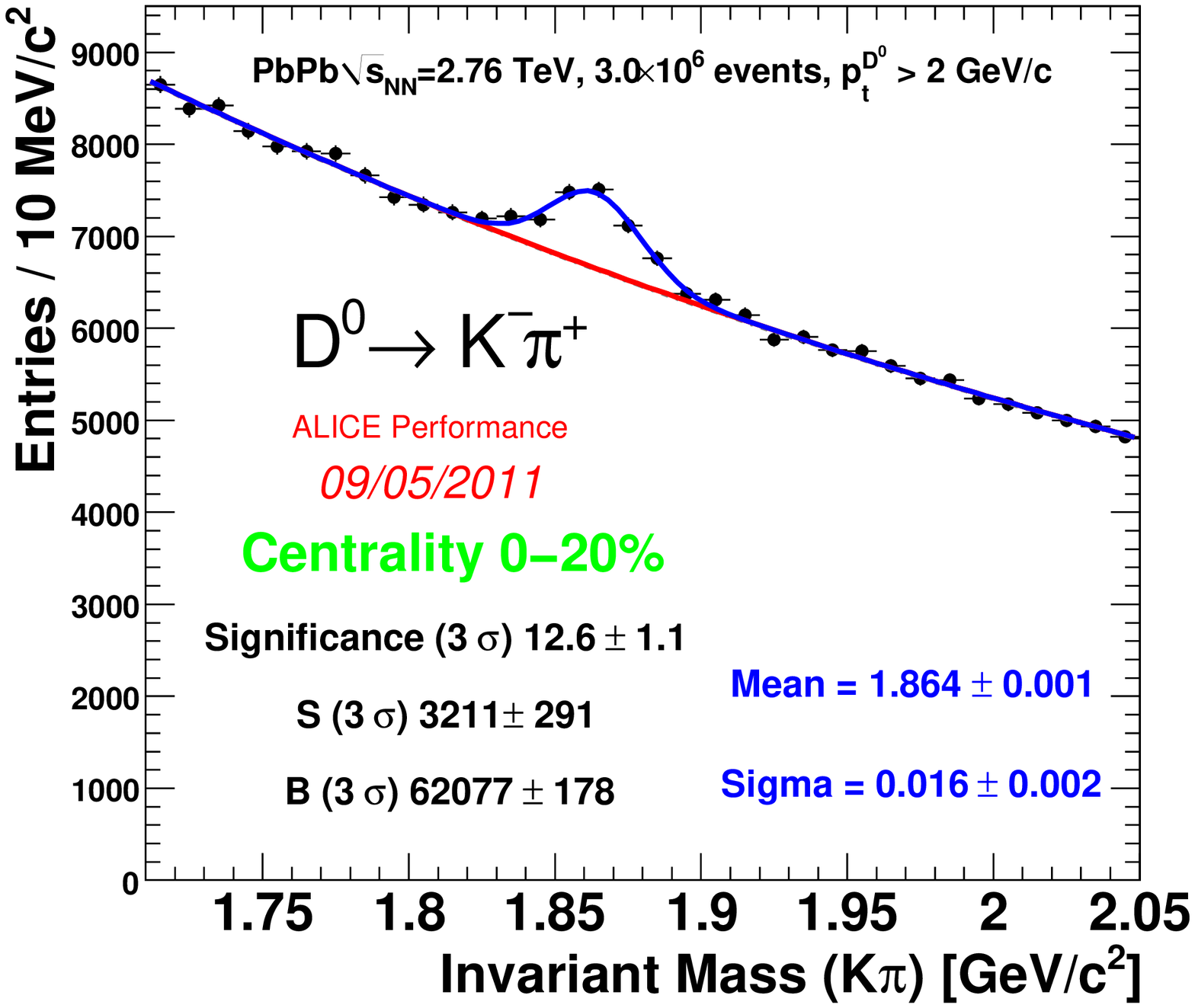}
\includegraphics[width=0.32\textwidth]{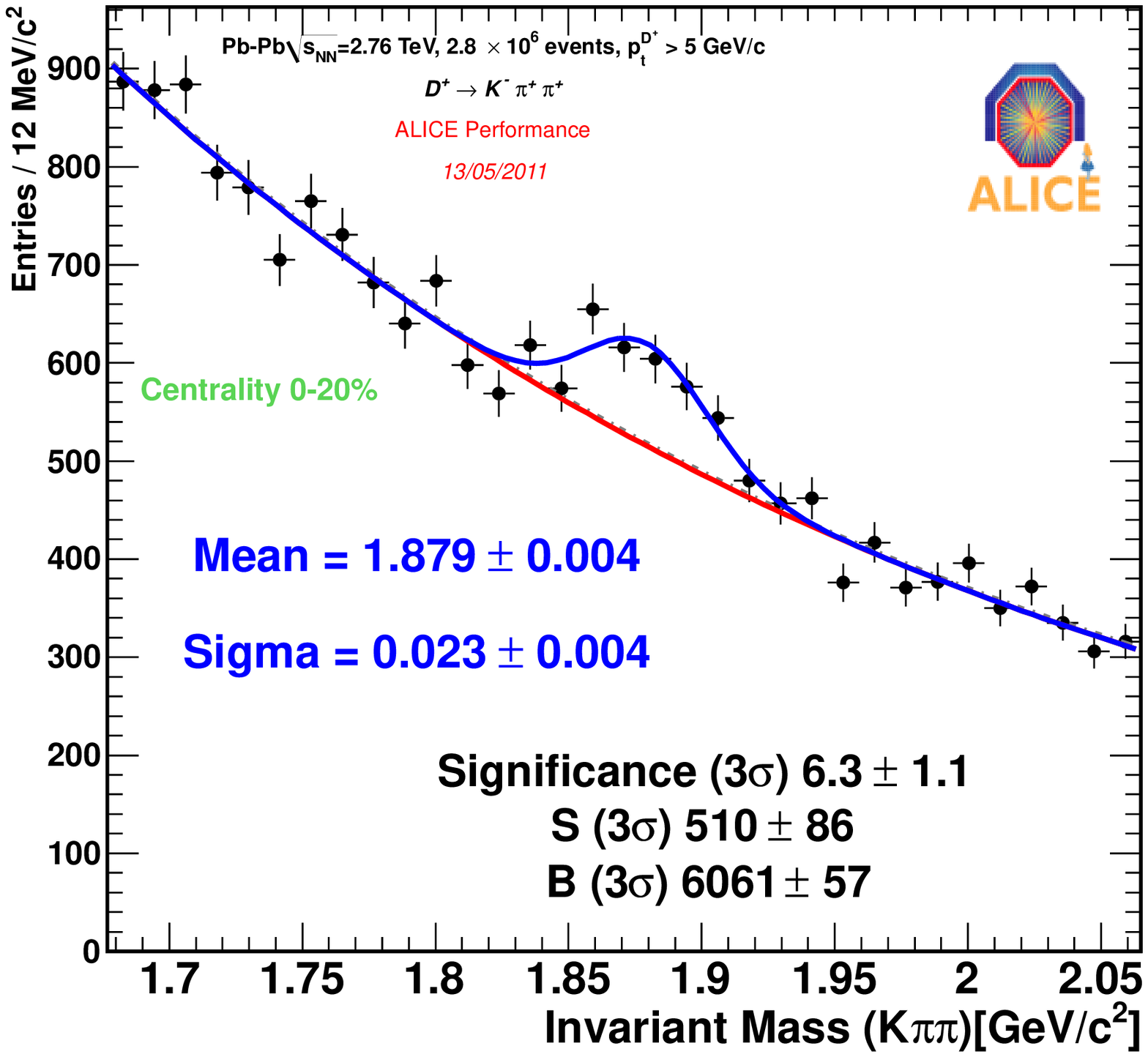}
\includegraphics[width=0.34\textwidth]{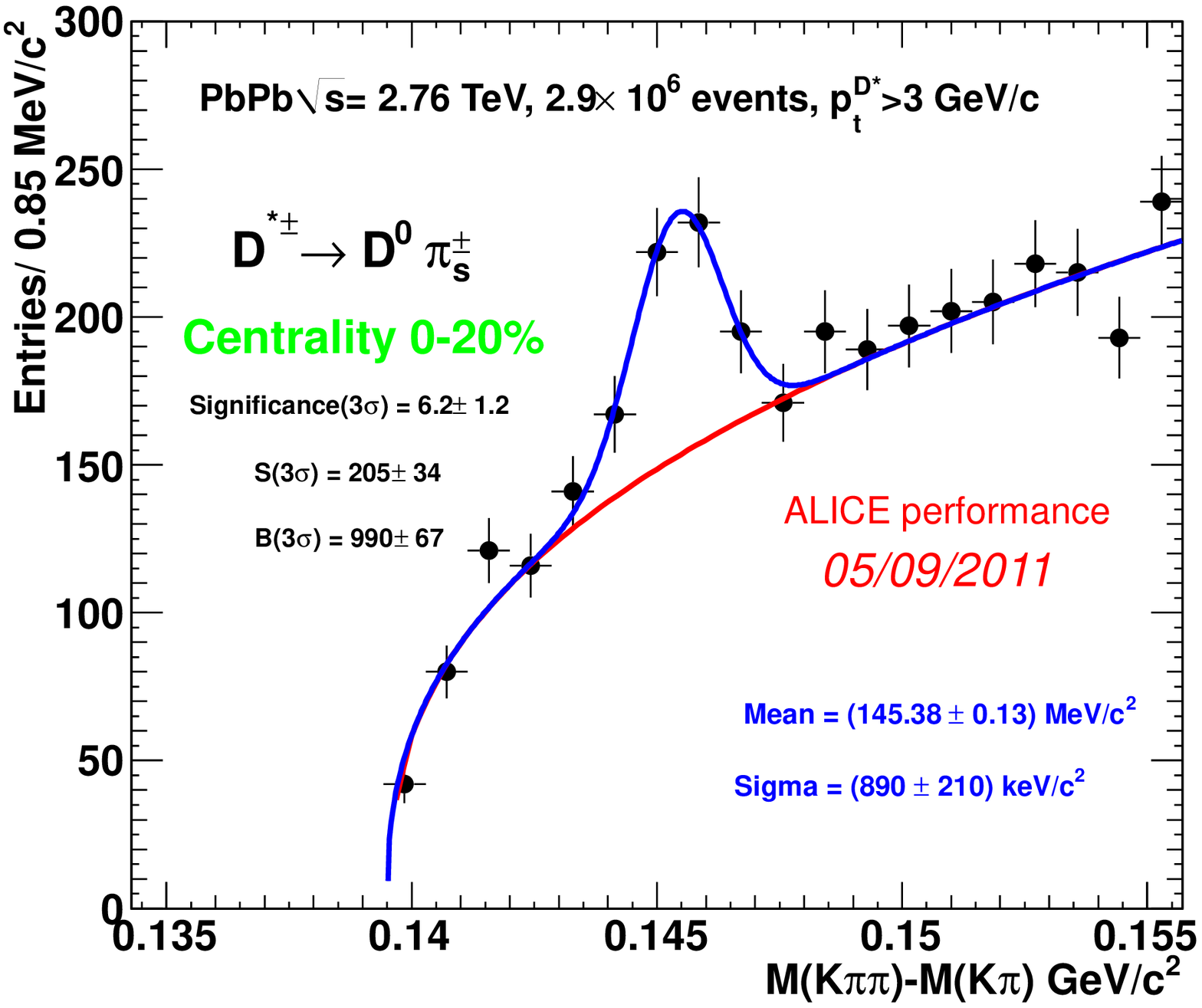}
\end{center}
\caption{D$^0$ (left), D$^+$ (center) and D$^{*+}$ invariant mass analysis in the centrality range $0-20\%$} 
\label{mass}
\end{figure}
The selection of the D$^0$ and D$^+$ decays (with mean proper decay length of $c\tau\approx 123$ and $312~\mu m$, respectively) was based on the reconstruction of secondary vertex topologies~\cite{Dmes}. In the case of the D$^{*+}$ decay, since the decay proceeds via strong interaction, the secondary vertex topology of the produced D$^0$ was reconstructed.  The cuts were optimized in intervals of $p_t$.
To further suppress the combinatorial background, a particle identification (PID) selection based on the specific energy loss in the TPC and on the time-of-flight  information provided by the TOF detector was developed. This selection provides a strong reduction of the combinatorial background in the low-$p_t$ region, while
preserving most of the D$^0$ and D$^+$ signal ($\sim 100\%$). In the D$^{*+}$ case, for the centrality class 0--20$\%$, in order to cope with the large combinatorial background, a tighter PID cut of $2\sigma$ level in the TPC was applied.

\section{Production cross section of D mesons in pp collisions}

For each D meson species the production cross section, measured by ALICE at $\sqrt{s}$ = 7 TeV~\cite{gianmic} and with an integrated luminosity of $1.4$ nb$^{-1}$, was scaled down to $\sqrt{s}=$ 2.76 TeV to be used as reference for $R^{\rm D}_{{\rm AA}}$. The theoretical $\sqrt{s}$-scaling~\cite{scaling} factors were defined as the ratio of the cross sections from the FONLL~\cite{fonll} pQCD calculation at 2.76 and 7 TeV. \\
\begin{figure}[h!]
\begin{center}
\includegraphics[width=0.45\textwidth]{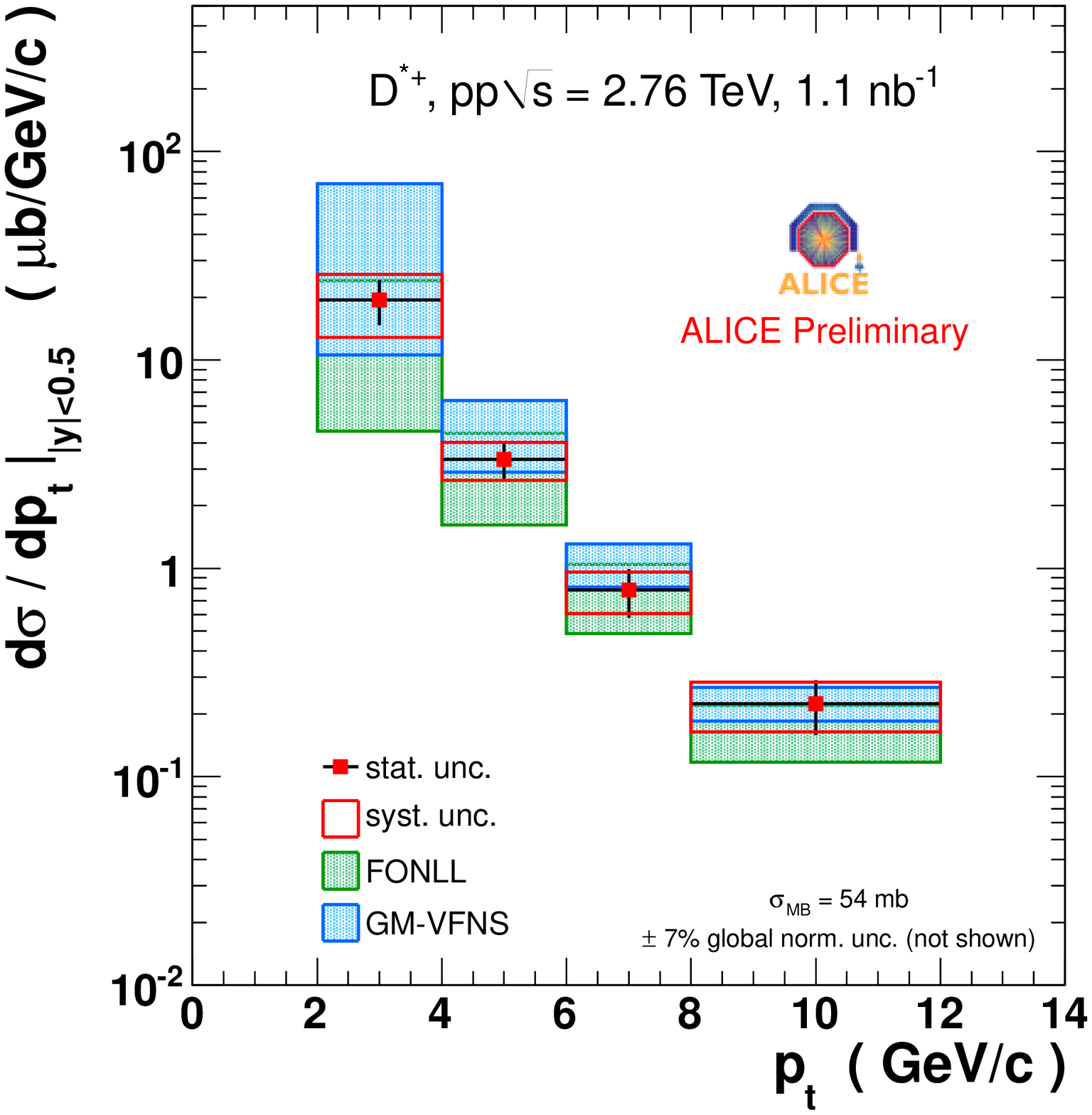}
\includegraphics[width=0.45\textwidth]{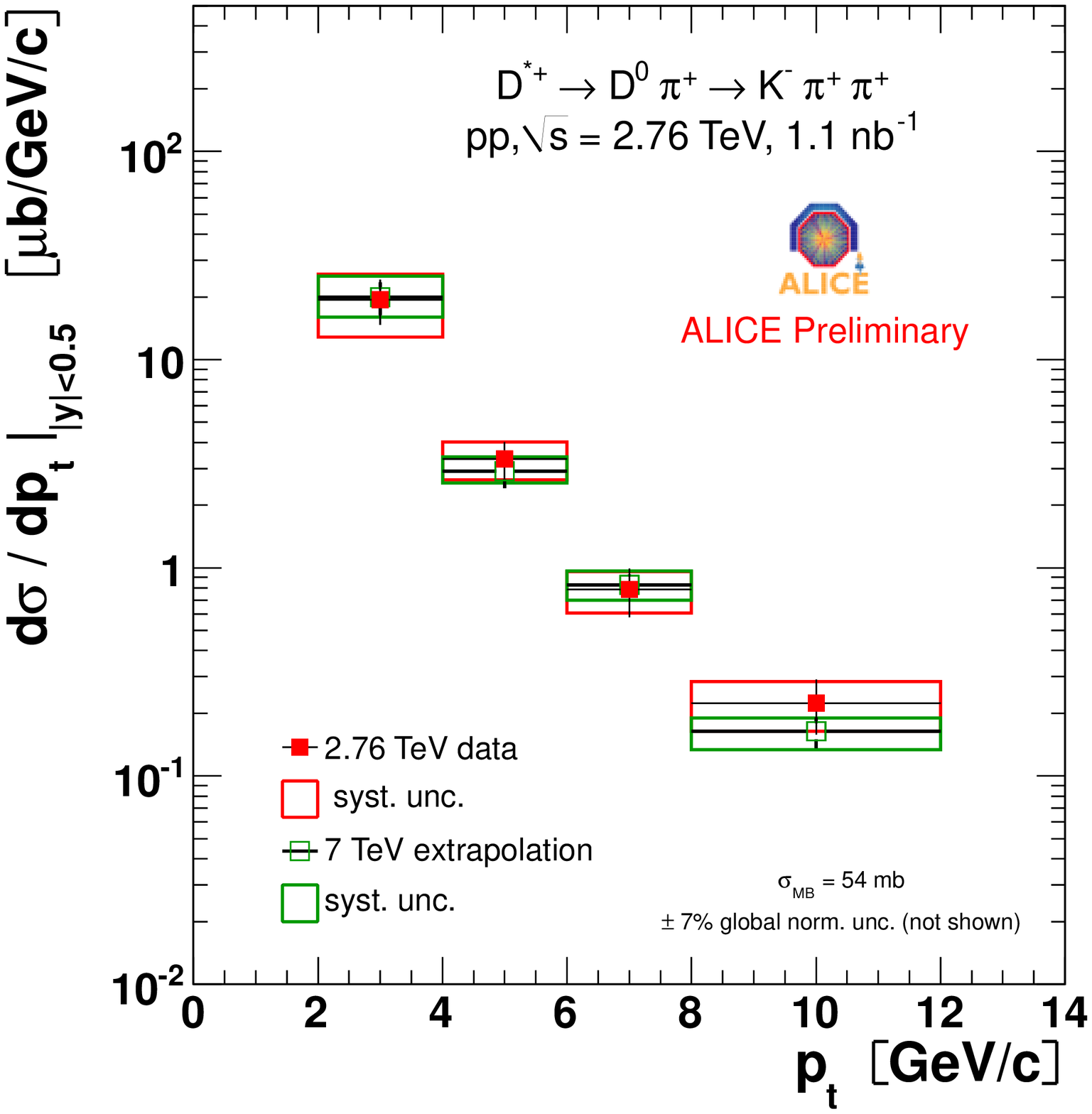}
\end{center}
\caption{Left panel: $D^{*+}$ production cross section measured at $\sqrt{s} =$ 2.76 TeV compared with FONLL and GM-VFNS \cite{VFNSalice}. Right panel:  Same plot compared with the theoretical scaling of the $D^{*+}$ production cross section at $\sqrt{s} =$ 7 TeV. Systematic and statistical errors are included.} 
\label{comp}
\end{figure}
The scaling uncertainty ranges from 25$\%$ to 10$\%$ from low to high $p_t$. A consistency check of the reference was done by comparing the result of the scaling with the D meson production cross section obtained with the low statistics pp run at $\sqrt{s} =$ 2.76 TeV (L$_{INT} $ = 1.1 nb$^{-1}$).
The $D^{*+}$ production cross section at $\sqrt{s} =$ 2.76 TeV as well as the result of the comparison with the theoretical scaling are shown in figure \ref{comp}. 

\section{D mesons nuclear modification factor}

\begin{figure}[h!]
\begin{center}
\includegraphics[width=0.49\textwidth]{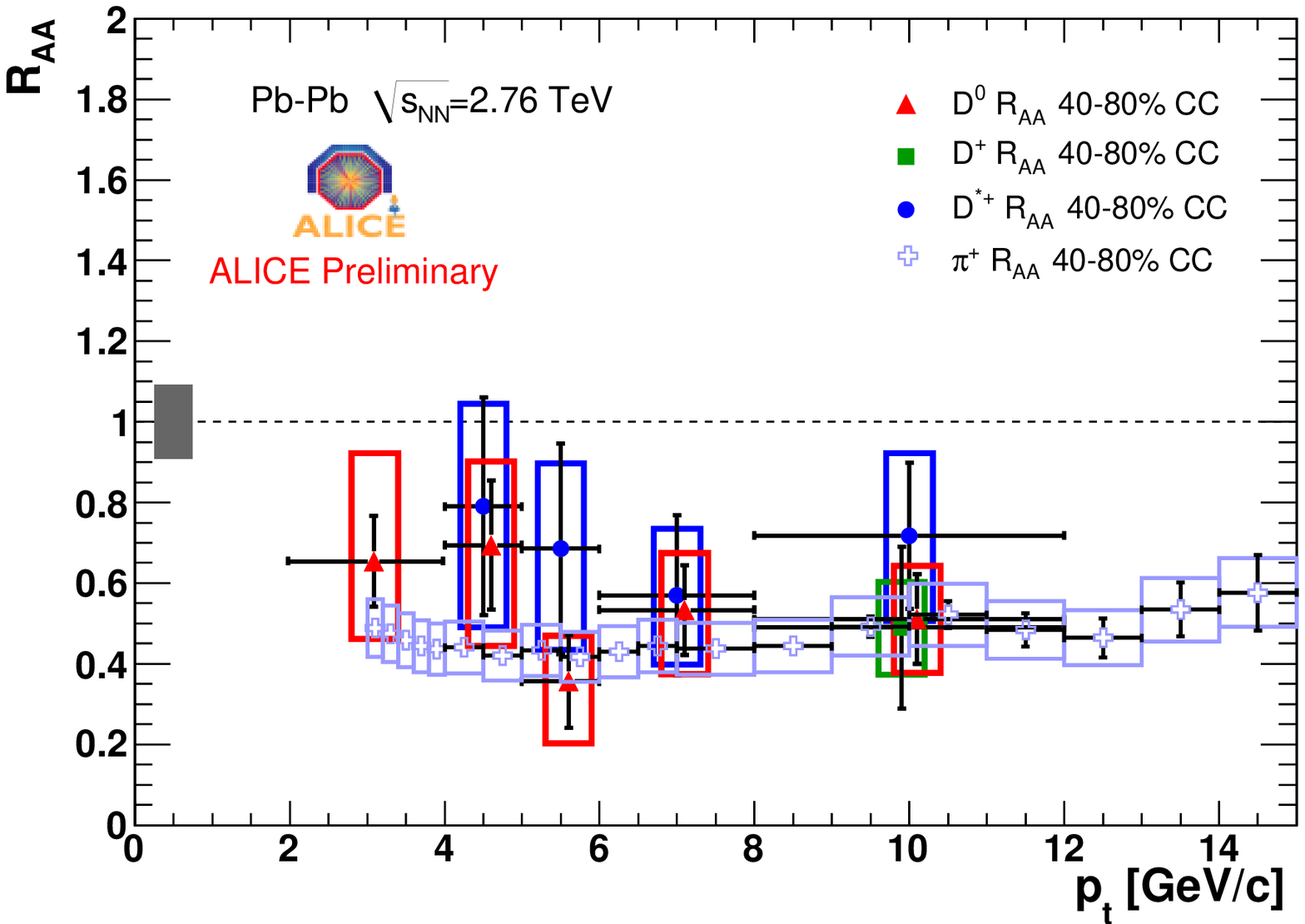}
\includegraphics[width=0.49\textwidth]{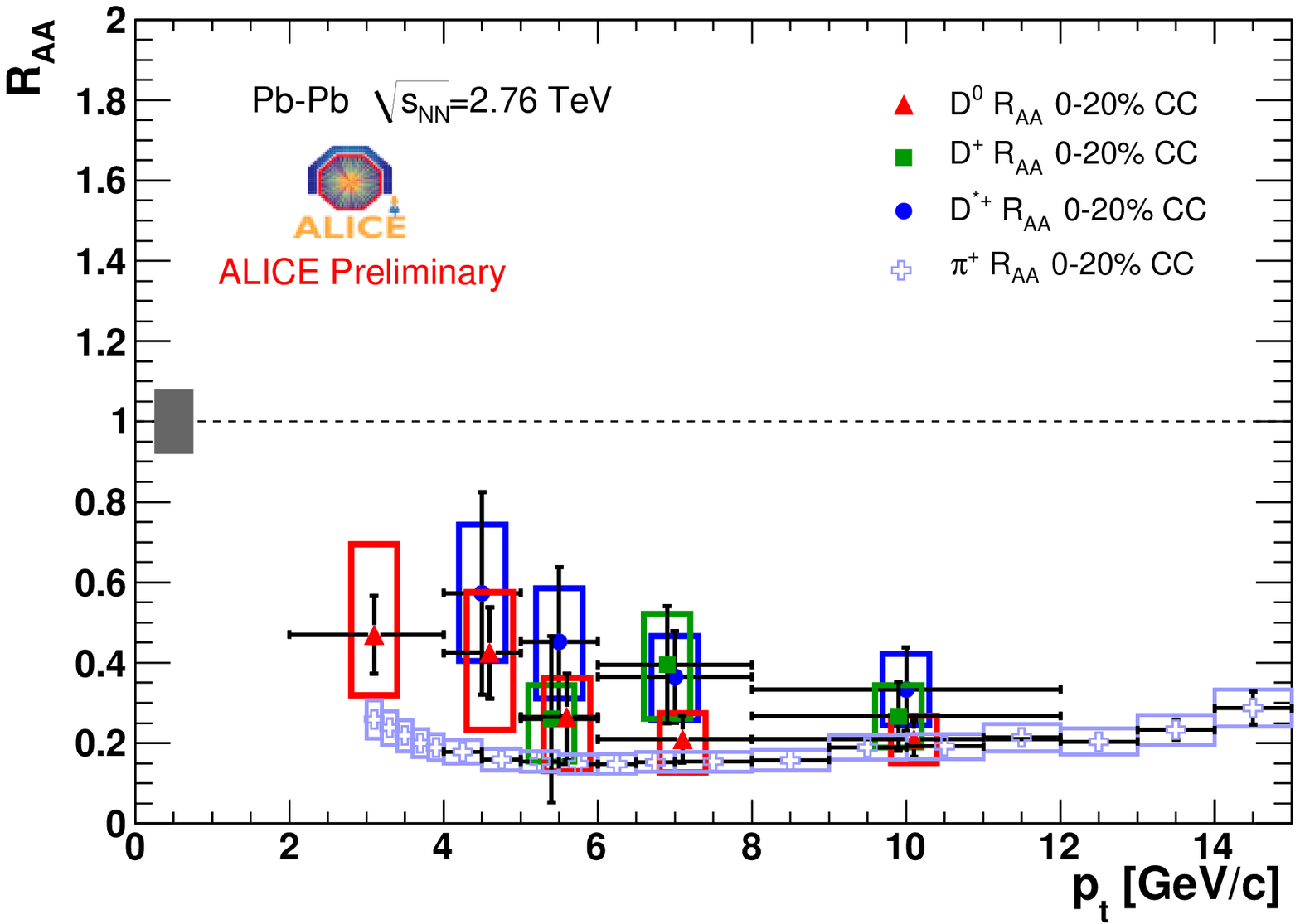}
\end{center}
\caption{D meson nuclear modification factor in the centrality classes $40-80\%$ (left panel) and $0-20\%$ (right panel). In both centrality classes the $R^{\pi^+}_{{\rm AA}}$ is shown for comparison.} 
\label{dmes}
\end{figure}

\begin{figure}[h!]
\begin{center}
\includegraphics[width=0.72\textwidth]{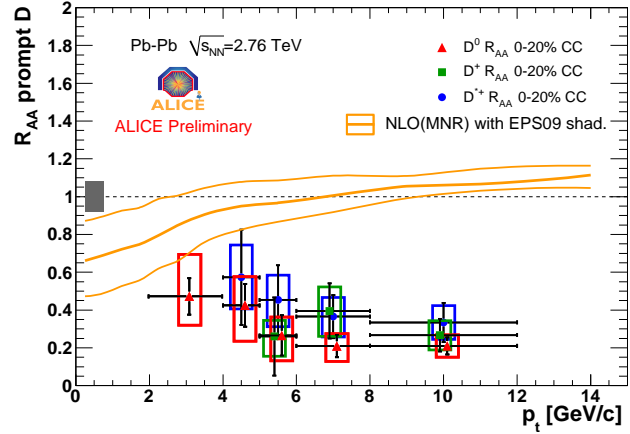}
\end{center}
\caption{Comparison of measured D meson $R_{AA}$ with shadowing predictions using EPS09~\cite{vogt}.} 
\label{dmes2}
\end{figure}

The nuclear modification factor of D$^0$, D$^+$ and D$^{*+}$ mesons is shown in figure \ref{dmes} for the centrality classes $0-20\%$ and $40-80\%$. The  $R^{\rm D}_{{\rm AA}}$ of the three D mesons agrees within uncertanties and  shows, in the most central class, a clear suppression of about factor 4-5 for $p_t >$ 5 GeV/$c$. The statistical precision of the mesurement is limited by the size of the 2010 Pb--Pb data sample to $\sim 20-25\% $ in the case of D$^0$ and to $\sim 30-35\% $ for D$^{*+}$ and D$^+$.
The estimated total systematic uncertainty, accounting for the uncertainties on signal extraction procedure, PID selection strategy, track reconstruction efficiency, cut stability and $R^{\rm B}_{{\rm AA}}$ hypotesis (hypothesis on B mesons $R_{AA}$ encoding all the potential nuclear and medium effects affecting B production~\cite{Ar}) varies from ~50$\%$ to 25$\%$ from low to high $p_t$.
The comparison with charged pions~\cite{AA} shows a hint of $R^{D}_{AA}>R^{\pi^+}_{AA}$ even if, with the current statistical uncertainty stronger statement cannot be made. In figure \ref{dmes2}  our measurement is compared with the shadowing expectation using the EPS09 parametrization~\cite{vogt}.   

\section{Conclusions}
We have presented the first measurement of the D$^{*+}$ nuclear modification factor and we  reviewed the status of measurement of the D$^0$ and D$^+$ $R_{{\rm AA}}$. The large suppression, of about factor 4, found in the most central class analyzed ($0-20\%$), shows the presence of the hot and dense QCD medium formed in ultra-relativistic heavy ion collisions.
The comparison of $R^{\rm D}_{{\rm AA}}$ with the one of charged pions shows a hint of  $R^{\rm D}_{{\rm AA}}>R^{\pi^+}_{{\rm AA}}$.  The results from the 2011 Pb-Pb run should allow for a more conclusive comparison.

\end{document}